\documentstyle[psfig,epsf,conf-X,10pt]{article}
\begin{document} 
\small
\heading{%
%
Large-Scale Structure in the \newline ROSAT North Ecliptic Pole Survey
}
\par\medskip\noindent
\author{%
Christopher R.\ Mullis$^{1}$
}
\address{%
Institute for Astronomy, University of Hawaii, 2680 Woodlawn Drive, Honolulu, HI, 96822 USA.
}

\begin{abstract}
We have used the ROSAT All-Sky Survey around the North Ecliptic Pole
to construct a complete sample of galaxy clusters.  The deep and
contiguous nature of the survey affords us the opportunity to examine
large-scale structure in the Universe on scales of hundreds of
megaparsecs.  We have identified over 99\% of the 446 X-ray sources in
the survey area.  The cluster sample consists of 65 objects with
redshifts approaching unity.  Surprisingly, some 20\% of the clusters
exists in a wall-like structure at z=0.088 spanning the entire
9$^{\circ}$ $\times$ 9$^{\circ}$ survey region.  This is a very
significant extension of both the membership and the spatial extent to
a known supercluster in this location.

\end{abstract}
\section{Introduction}
Within the ROSAT All-Sky Survey (RASS) \cite{Voges}, the region around
the North Ecliptic Pole (NEP) is special because the exposure here is
deepest where the great circle scans overlap.  Furthermore, galactic
obscuration is not severe here at a galactic latitude of +30$^{\circ}$.
Note the South Ecliptic Pole exposure depth is degraded by the South
Atlantic Anomaly and the extragalactic sky in this direction is
partially obscured by the Magellanic Clouds.

Over the last nine years we have executed a systematic optical/NIR
followup program of imaging and spectroscopy primarily at the Mauna
Kea observatories.  We have identified the physical nature of 99.3\%
of the 446 X-ray sources detected at greater than 4 sigma in the
9$^{\circ}$ $\times$ 9$^{\circ}$ survey region.  The initial observing phase
of this project is now complete, and our emphasis is shifting to
analyzing the statistically complete, X-ray selected samples.

\section{The Cluster Sample}
We have constructed the NEP cluster sample to examine X-ray cluster
luminosity evolution and to characterize large-scale structure (LSS).
The initial evolution results are presented by I.\ M.\ Gioia in these
proceedings.  Here we present the preliminary findings on LSS.

We emphasize the cluster sample is unique in that it is both deep and
contiguous.  The sample consists of 65 objects discovered in a 81
square-degree region with a median sensitivity reaching 8 $\times$
10$^{-14}$ erg s$^{-1}$ cm$^{-2}$.

The observed NEP cluster redshift distribution shows a striking
feature at z=0.088.  Fifteen clusters lie in the redshift interval
0.07 to 0.10, an impressive 23\% of the entire sample.  This interval
is 4 to 5 times more populated than expected which is significant at
the 4 sigma level. The complex spans the entire survey area.

\begin{figure}
\centerline{\vbox{
\psfig{figure=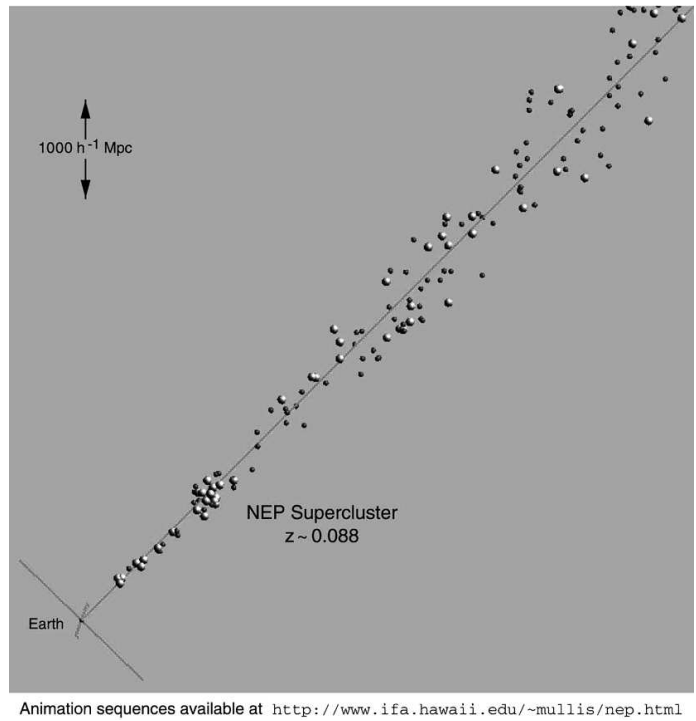,height=9.5cm}
}}
\caption[]{3D view of the near section of the conical
NEP survey volume.  The Earth is at the origin with the long, redshift axis
through the NEP.  Bright, large spheres are clusters and groups of
galaxies.  Small, dark spheres are AGN.}
\end{figure}

\section{The NEP Supercluster}

LSS has been previously seen in the direction of the NEP.  Fourteen
years ago Batuski \& Burns \cite{BB} produced a finding list of candidate
superclusters based on a percolation analysis of Abell clusters.  They
found an association of 6 Abell clusters approximately 5$^{\circ}$ from
the NEP.  Subsequently Burg et al.\ \cite{Burg} reported 5 X-ray cluster
candidates detected in an early ROSAT pointed observation possibly
related to this supercluster.

IRAS mapped the sky in a fashion similar to ROSAT and hence has the
same super-sensitivy at the NEP.  Ashby et al.\ \cite{Ashby}
endevoured to test starburst galaxy evolution with a sample of IRAS
NEP galaxies but instead were overwhelmed by LSS.  Fifteen galaxies,
20\% of their IR selected sample, turned up at z=0.088.  Rinehart et
al.\ \cite{Rinehart} have expanded the IRAS survey to the same area on the
sky as our survey, and they continue to see a ``massive'' sheet of
galaxies, estimated to be 3 times as dense as the Great Wall.  A final
signature of structure in the NEP region is a low-z Lya absorber
detected at z=0.089 in the direction of a well-known QSO H1821+643
from Tripp et al.\ \cite{Tripp}.

We have prepared an animated, three-dimensional ``fly-through'' of
the NEP survey volume which is available over the web at \linebreak {\em http://www.ifa.hawaii.edu/\~{ }mullis/nep.html}.  Fig.\ 1 presents a
demonstrative frame from the animation.  Thanks to the ROSAT NEP
Survey we have a much improved understanding of the NEP supercluster.
We have more than doubled the physical size of the original Batuski \&
Burns supercluster and tripled the cluster content.  We now know it to
consist of at least 20 clusters and groups and 12 NEP-detected AGN.
It has a planar distribution of 70 $\times$ 70 $\times$ 25 h$^{-1}$ Mpc
with a 12$^{\circ}$ extent on the sky.  

The NEP structure compares favorably in size and content relative to
well-known objects such as the Great Wall and the Shapley
Supercluster.  It is not the largest nor the most dense but one of the
most robustly sampled.  Twenty X-ray emitting clusters of galaxies
demark the highest density regions while the over sixty IRAS galaxies
and AGN trace out the lower density domains.  Furthermore, the door
is open for the supercluster to be even larger;  its depth and
thickness are well-constrained but its breadth is currently not bound.
We have only examined one side of the original NEP supercluster.

\section{LSS at the NEP - Work in Progress}

We are currently working to chart the true extent of the NEP
supercluster by discovering additional galaxy clusters beyond the
original NEP survey region.  Optical and X-ray followup observations
will be used to examine the dynamical state of the structure.  Is it,
or any portion of it, gravitationally bound?  Are alignments between
clusters detectable in their X-ray contours?  The supercluster's
edge-on aspect makes it an attractive candidate in which to search for
X-ray filaments.

Here we have largely concentrated on the NEP supercluster.  But the
rest of the cluster sample should not go unscrutinized.  Lesser
examples of clusters of clusters, potential superclusters, are visible
in three-dimensional examinations. A correlation function analysis
will quantify the degree of clustering present.

The NEP AGN sample of 211 objects is another means for examining LSS.
Though they are more sporadic markers compared to clusters, they reach
to greater depths (z$_{max}\sim$4).  Our sizable collection is free
of the serious selection effects that plague non-X-ray selected
samples.  There are several instances of clusters of AGN visible in
three-dimensions.  The global clustering characteristic of the AGN
will be determined via a correlation analysis.

\section{Summary}

Experimental design and serendipity have converged at the NEP to
reveal a remarkable element of LSS in the X-ray Universe.
The NEP supercluster is very large and likely even larger.  A better reckoning of its girth and content have been revealed by X-rays.
There are more LSS results to come from the NEP cluster and AGN samples.
Finally and perhaps most importantly, the ROSAT NEP survey demonstrates the
effectiveness of X-rays for studying LSS and opens the door for future
X-ray survey missions to pursue the subject further.

Details of the NEP supercluster discovery in the ROSAT NEP survey will
be presented in a forthcoming paper by Mullis et al.\ \cite{Mullis}.

\acknowledgements{The ROSAT NEP survey work is the subject of the
author's PhD thesis being completed under the supervision of Pat
Henry, and is a collaboration with Isabella Gioia, Hans B\"ohringer,
Ulrich Briel, Wolfgang Voges, and John Huchra.  Partial financial
support comes from NASA grant S99-GSRP-019.  We wish to thank Manolis
Plionis, Ioannis Georgantopoulos and the other LOC members for
organizing such an intensive and enriching meeting in such a memorable
locale.}

\vfill
\end{document}